# On the Emerging Area of Biocybersecurity and Relevant Considerations

Xavier-Lewis Palmer[1], Lucas Potter[1], and Saltuk Karahan[(2)]

[1] Biomedical Engineering Institute, Department of Electrical and Computer Engineering, Old Dominion University, Norfolk, VA 23529, USA

[2] Department of Political Science and Geography, Old Dominion University, Norfolk, VA 23529, USA

skarahan@odu.edu

**Abstract.** Biocybersecurity is a novel space for the 21st century that meets our innovations in biotechnology and computing head on. Within this space, many considerations are open for and demand consideration as groups endeavor to develop products and policies that adequately ensure asset management and protection. Herein, simplified and brief exploration is given followed by some surface discussion of impacts. These impacts concern the end user, ethical and legal considerations, international proceedings, business, and limitations. It is hoped that this will be helpful in future considerations towards biocybersecurity policy developments and implementations.

## 1   Introduction

Biocybersecurity presents a new way of exploring how we protect our societies. It can be thought of in part as an extension of cybersecurity, which involves the protection of systems, made of hardware and software, from unauthorized access and attacks. Biocybersecurity is alternatively referred to as Cyberbiosecurity, which, according to the Peccoud Lab, exists at the intersection of cybersecurity, cyber-physical security, and biosecurity, and focuses on mitigating risks within and relating to their intersections [1, 2]. A growing need exists for expertise in this field as we persist in a world at a time where computer systems and biotechnology are increasingly ingrained in day-to-day life, in both developed and developing economies. Furthermore, strong lines eventually need to be drawn when determining where Biocybersecurity and other fields end, to adequately allot resources and focus towards adequately mapping vulnerabilities and prevent exploits that may occur and evolve [1].To this end, we have provisionally defined Biocybersecurity as thus: Any cybersecurity system where a biological component, target, or interlock is involved in the terminal or intermediate stages.

The first potential hurdle- is there a clear and present need for the development of an entirely new field of study? Let us first turn to the current ease of obtaining biological data. Whereas sequencing DNA used to be a rigorous process, it has gotten easier[3].In fact the Human Genome Project was projected to last 15 years, and was completed in 13, demonstrating that the speed of computing and insights into the structure of life have made it easier than ever to obtain, disseminate, and utilize biological data [3]. Secondly- the processing and accessibility to the mentioned data is easier than ever. The physical barrier to acquiring healthcare data has been demolished in the name of ease of access and patient-centered care digitally. So, while the hardwaretocrackconventionalcybersecuritybarriersismoreprevalent,thesafeguards of that data have been chipped away. Thirdly, new computational platforms question the nature of the separation of biology and computing, leading to a more tightly integrated biocybersecurity process [3, 4]. Some rising platforms even call into question the contemporary understanding of typical cybersecurity processes and could circumvent typical security at the cost of creating entirely new (and unforeseen) problems that arise from biological matter and the application of medicine being used to perform computations [1, 4]. An entire new group of subdisciplines may be needed to understand the unknown complications that arise from use of these platforms [5]. All these features of modern healthcare and technology combine, meaning, that biological data can be applied in more ways. For instance, the process of implicit authentication using biological data is now possible with COTS(Commercial, Off-The-Shelf) components. A smartwatch can access multiple kinds of information including heart rate data and more [6]. The matching a set of recorded data to a user accessing a work terminal. However, this means that said data would, somehow, be accessible to other, perhaps nefarious, individuals. As touched on above, part of the rise of potential threats to in the field of biocybersecurity is the demand for easier and faster access to data. Patients desire faster, more convenient access to medical records; medical research companies need larger and more comprehensive trials; and data sets to remain viable for the more technically demanding medical interventions currently used [7–10]. Medical companies also need to increase their awareness of the potential of malware to compromise device outputs; for example, one research team's recently demonstrated an algorithm that could modify CT scans to mislead sick patients into believing that they are healthy and vice versa [11]. Inadequate defenses against such malware and inadequate protection of patient data could brew a maelstrom of related crises, on the scale of ransomware outbreaks [11]. The prevalence of such devices that demonstrate the confluence of biology and cybersecurity include thumbprint scanners, retina scanners, digitized healthcare records, forensics databases, DNA sequencing databases, and pharmacology records. All of these could be accessed and used for threats in the Biocybersecurity domain. The potential growth of biological data in demand is seen in examples in the rise of services to sequence and interpret DNA as seen through ancestry and health services through DNA analysis. For a more concrete example of what is meant when they describe the definition of biocybersecurity: let's say that a user is at a computer using implicit authentication. The computer's security system tracks the eyes of the user, and if the saccic rhythms change, the computer locks the user. Under our definition, this would fall under biocybersecurity, as the system is using biological inputs or data as an intermediate step- in this case as preventative interlocks.

## 2  Impacts and Considerations

### 2.1  Some Ethical Considerations

Let us consider some ethical conundrums that help us alternatively view problems within the field that relate to the previous examples. We'll do so with a few, including the trolley problem and the Gettier Problem. The trolley problem states that a trolley can head down two tracks, and a moral operator selects which track to send it down- killing whoever is on either track. One considerable problem in this context is that with many of these bulk DNA or other -omics analyses, the operator is not directly at the switch. In fact, they may produce or remove multiple switches that complicate the ethical calculus involved with handling biological data that is digitalized. The use can get out of hand. The operators' data collection gives way toward many more switch pullers, those who get their hands on the data, who can affect a great many lives beyond their original intention. Now let us pivot to another potential scenario with problems that might be faced: a business has access to a large amount of personal DNA data that gets mis-used. They use it to discover the prevalence of people that enjoy a certain kind of sugar more or consume considerably more than others. If they then mold their advertisements based on this data- either personally with internet-based ads or in bulk- are they responsible for the health of this population of people? With this data, companies within the food and beverage industry have an extra curated set of people who may have a higher difficulty of exercising proper agency over their dietary decisions. At the same time, they are working against their interests to not use such data. Insurance agencies might obtain and use this data with regards to modifying payment or coverage rates with diseases such as diabetes and other metabolism linked disorders in mind, their business model and consumer finances to juggle. Afterall, this must be viewed additionally in the light of emergent actions that appear from this data merely being available and obtainable. Once produced, it is subject to analysis by actors within who may not share company motives, companies that have data sharing agreements, malicious actors that may leak or funnel said data to other groups at multiple levels of power, and even harder to discern, meta- analysis by an independent company who is able to link said DNA to the customers involved. Furthermore, analyses, both core and meta, of this data by companies that they share this with can lead to further emergent concerns of abuse. To summarize the Gettier Problem, it is a case on which an agent can have a justified basis for belief in a proposition and still be wrong about the fact of a matter. The Gettier Problem can be used to prompt us, within biocybersecurity, to consider what data we collect and hold, but also ask why and in what form and under what conditions we interface with the data and meta-data in terms of the risks posed if we are wrong about how such data might be used and abused. One must bet on the possibility of being wrong and the consequences that follow from that. We can relate this to biocybersecurity in which the same company above has a justified belief in their level of security, and yet suffers a breach that results in the leak of millions of sensitive and valuable biometrics. Companies must be willing to maintain ethical boards and rigorous standards to limit unnecessary data collection, holdings, transfers, eyes on data, and time of holding said data. They must be prepared for not only damage control, but compensation and talks with the public that they serve so as not to damage perceptions of biocybertechnologies with the public. For example, is the user of a hypothetical system made aware that their eye movements or other biosignatures, or biological means of expression are being tracked? Can the user manually turn this feature on and off? If no they should be made aware and be given the ability in case of potential abuse that they may have overlooked, the company must be ready to responsibly deal with the mountain of problems that may follow. In general, ethical approaches to biocybersecurity must be comprehensive.

2.2 End User and Social Impacts

As Biocybersecurity policy matures, it is ever prudent to consider social implications that exist in a world where biocybertechnologies and those adjacent become more prominent and their misuses become more of a threat. This increasingly applies to interconnected technologies that we easily and often take for granted, especially those that have benefitted from recent life sciences research as mentioned earlier. Let us consider the "Internet of Things", known as IOT and the devices that can fall under this paradigm. IOT can be thought of as a mass constellation of devices connected to the internet [12]. You may recognize them in the form of commonly used products such as refrigerators that report on the quality of its contents or remind you on when to restock, your wearable exercise equipment that gives your heartrate or temperature, medical autoinjectors that monitor or regulate your insulin supply and report to your doctor, rooms that monitor your position and try to keep the room at a suitable temperature for you, implants that augment features of your body, or even more simply, your smartphone, with its bevy of sensors. Each of these devices gather and transmit a variety of data that can directly or indirectly characterize consumers in ways that they may or may not consent to. Quite easily, a consumer can consent to the use of a device that monitors their heartrate, but to a skilled analyst, studying the heartrate over significant amounts of time can reveal one's sleep, work, schooling, romantic, diet, and social behavior, in ways that the consumer certainly wouldn't easily consent to. The same considerations can be applied to the earlier mentioned refrigerator, exercise equipment, and medical equipment – they all can give data which can generate a mesh of complex stories in different curated combinations and when interpreted differently [12]. Even when guarded with a degree of caution, a skilled hacker can gain access to said data, which leaves an ever-existing risk with the nature of said technology for anyone with privacy in mind. With such data ever open to exploitation, one huge social implication is increased societal fear, and one to follow is the erosion of trust in advanced technology. Depending on how ingrained said technology is within companies or governments, this can mean erosion of trust and cooperation with those entities as these technologies become increasingly exploited at the detriment of citizens. Companies and large governments would do well to tread cautiously while and when employing these technologies. Failure to reign in control of said data could lead to mass social disarray, which could be irreparably injurious to societal stability, depending on the extent of the damage. One more source of lay perceptions of biocybersecurity that no doubt affect policy is popular media in how it has influenced how we may see and interact with such technologies. One considerable influencer is that of Cyberpunk culture, which encompasses futures that push the boundaries of technologies, leading to a blending or enhancement of humans and their technology in often unique and beneficial ways [13, 14]. In some stories, innovations within such literature often arise from lack the of oversight, or reduced confidence in the ability of the government to adequately assuage needs of a growingly frantic populace in the face of ever-growing technological reliance. Cyberpunk culture has also contributed to the growth of cultures like that of the Maker-movement, which is composed of individuals that are often resisting traditional, institutional control of technologies while self-policing [13–15]. With respect to biology, some of them are addressing prosthetics, implantable electronics, gene editing and protein engineering, and bio-adjacent biotechnologies with wide appeal and potential to correct for deficiencies in their communities [14, 15]. An easy case to consider is that of the failure of the US government to control for drug prices, leading to some groups to take matters in their own hands to make them and analogues themselves such as some community bio labs that have met, with increasing success at mobilizing the community [13–15]. Some helpful efforts have just been to engineer other means of producing food, whereas others may aim to re-write some parts of life itself through the possibility of creating synthetic organisms [13–16]. Some of these groups may or may not apply for government funding and instead pursue their own path to innovation through private or self-funded measures. Examples can be seen among a few groups in the Community Bio movement which arose out of the Maker Movement, in which groups of people have been inspired to pursue these research areas and more through a mix of traditional and non-traditional cooperation with mixed success [17– 19]. Plenty of these successes resulted in the creation of start-ups that deal in a great amount of biometric data or material which is tracked, for improving health outcomes or expanding functions [14–19]. Much of what is thought of as cyberpunk in science fiction has reached reality, and this implies that the time to think ahead regarding the protection of biometric data is now. There's little reason to suggest that these projects won't become even more complex. Overall, there is much to consider socially as we consider and pursue cyberbiotechnological policies that address our increasing reliance and potential overexposure to such technology. Given that the base technology already exists by large and data is already being generated in volumes and at rates at which already has the potential to overly stir and stoke negative public action, we are able to need to bolster our foci on further social implications of such technology. To fail to do so can undo many societal gains within technologically advanced nations.

## 2.3    Policy and Legal Impacts

Quite a few policies have provisions and objectives that would be wise for groups to consider factoring into their cybersecurity policies. Some worthy of mention are the Nagoya Protocol, the Genetic Information Nondiscrimination Act, Dual Use Research of Concern, and the Health Insurance Portability and Accountability Act [2]. Each will be briefly summarized and drawn from. The Nagoya Protocol, known The Nagoya Protocol on Access to Genetic Resources and the Fair and Equitable Sharing of Benefits Arising from their Utilization (ABS) to the Convention on Biological Diversity, is an act sign in 2010 that discussed a framework for sensible sharing and access to genetic resources, in the scope of preserving biodiversity [2, 20]. Aims of it are ensuring, flexible, consent-based access to genetic resources that respect the jurisdiction to which said resources and or their owners belong to protect commercial and otherwise academic chains of value. The Genetic Information Nondiscrimination Act was passed in 2008 and has the aim of protecting people from DNA based health insurer and employee discrimination. A core weakness is that life and disability insurance as well as long term plans are not covered, leaving people up to the mercy of state government laws [2, 21]. Dual Use Research of Concern Policy, implemented in 2012, underlines policies to regulate life sciences research that could have a double edge. Means of research that pursue overly risky methods or are otherwise unethical face defunding and additional potential penalties [2, 22]. Lastly, the Health Insurance Portability and Accountability Act, originally enacted in 1996, outlined a means of protecting citizen medical data while making it available to health professionals in order to allow adequate, if not superior care [2, 23]. This is increasingly important in an industry where data driven care is used to deliver more informed treatment wherein health professionals can learn of complications and patient differences, allowing for a more personalized and accurate care of an individual, reducing confusion. What can be taken from the existence of these policies is that in the context of cybersecurity, biocybersecurity policies will need to flexible, based on consent, and the benefit of whose information is at risk. Policies not taking this into account are likely to face considerable legal action.

2.4 International Impacts

Historically, the connection between health and international security was based on the spread of the diseases and disease-related casualties in wars. As scholars focused on political stability and considered the relation between democracy, growth and political stability, the relation was mostly seen as a democracy supporting or hindering role in growth and stability. [24] However, democracies also require economic and social well-being for political stability. A deterioration in public health has the potential of impacting political instability and public unrest in democratic nations. [25] This direct relationship between public health and political stability allows international actors (mostly autocratic ones) to use public health as a tool of coercive power in international relations. One other factor influencing general public health is the ascent of globalization with its positive and negative influences. While globalization has been a factor for the spread of diseases, the technology that brought widespread connectivity has at the same time highly benefited the health service providers in reaching populations in remote corners of the world [26]. The management of resources and making them available in most needed areas were facilitated through global connectedness. The link between interconnectedness, management of resources and making them available in most needed areas were facilitated with the global connectedness. All of this naturally brings concerns about the security, reliability, and resilience of this connectedness. The link between interconnectedness, public health and political stability naturally make biocybersecurity a concern for international relations. Furthermore, this relevance is strengthened with the consideration of international political economy. Using the example from the Ethical Considerations section above, the fictious scenario below can help us understand the effects. Let's assume that one nation or NGO finds that a competing nation is particularly susceptible to a specific non-infectious condition. To make this example concrete, let's say Nation A is particularly susceptible to type 2 diabetes mellitus, and Nation B is a producer of Fructose. If Nation B increases marketing, increases supply, and seeks trade deals to increase the uptake of fructose in Nation A-does this count as a kind of stochastic act of war? This and similar questions form the landscape defining the impact of biocybersecurity on international security.

2.5   Business Impacts

Potential business impacts of biocybersecurity policies are wide reaching, leading significant effects of consumer trust, intellectual property (IP), domestic and international supply chains, and ultimately capital. In terms of consumer trust, the business impacts seen today from data breaches are likely not as far-reaching as will be true in the future. This is potentially due simply to the lack of understanding of the value of user data and comparatively reduced coupling of individuals to data. As the more technically literate people become the primary consumers of such business, such mishandling of data breaches may have more dire consequences- especially if the data is biological in nature. In the same vein, as the demand for faster innovation rises, more biological IP (potentially even in the form of Trade Secrets) will be placed in digital formats, which renders it vulnerable to theft. Additionally, the new avenues of security breaching will not solely be in the domain of more advanced technology. As historical examples demonstrate even the most current technologies can be breached by relatively simple methods [27, 28]. A tangentially related topic would be the targeting of the synthetic biology supply chain. The supply chain of any manufacturing company is an important matter, but the biological field has its own major dangers and limitations when considering the items that it requires and exports [28]. At the core of the problems above, considerable economic damage in the form of lost capital and weaknesses economic sectors linked to cyberbioeconomies is possible.

## 2.6 Complications and Limitations

The emerging spread of data processing methods into biological domains is, from a scientific perspective, a worthy and meaningful goal for many fields. However, the demand for faster and easier accessibility, the dependence on biometric data for security purposes, and the potential growth of automated biological analysis is a looming threat in the coming world. It is the hope of the authors that two goals were accomplished by reading this work. (1) That the "Failure of Imagination" that led to many threats in the cybersecurity domain will occur to a lesser extent after reading this work. (2) That the novelty of biodata will be seen as it is- not just an interesting benefit of the coordination of biology and computational sciences, but as an venue of attack.

## 2.7 Conclusion

There is little rationality in denying the currently robust and the potentially explosive growth of biocybersecurity as a field of thought, research, and action. The dangers presented in this paper, as well as the complications of our world are clear. The question is not one of what to do if these problems occur, but what do we do to prevent, ameliorate, and treat them? What are we going to do about it?